# Designing a Truly Integrated (Onsite and Online) Conference: Concept, Processes, Solutions

*By Alexei Botchkarev[1], Lian Zhao[2], Hamed Rasouli[2]*
*[1]Ministry of Health and Long-Term Care, Ontario*
*[2]Ryerson University, Ontario*

**ABSTRACT**

Web conferencing tools have entered the mainstream of business applications. Using web conferencing for IEEE conferences has a good potential of adding value to both organizers and participants. Authors propose a concept of Truly Integrated Conference (TIC) according to which a multi-point worldwide-distributed network of conference online authors/participants will enhance the standard (centralized) IEEE conference model, which requires attendance of the participants in person at the main conference location. The concept entails seamless integration of the onsite and online conference systems, including data/presentation, video, audio channels. Benefits and challenges of the TIC concept are analyzed. Requirements to the web conferencing system capable of supporting the TIC conference are presented and reviewed against commercial web conferencing tools. Case study of the IEEE Toronto International Conference – Science and Technology for Humanity, which was the first realization of TIC, is presented which analyzes various aspects (organizational, technological, and financial) of the integrated conference.

## 1.0 Introduction

IEEE is the single biggest organizer of conferences in the world sponsoring over 900 conferences annually. These events offer IEEE members and engineering community at large a great opportunity for professional communication and publishing their research results in the IEEE Xplore database - the world's highest quality collection of technical literature.

IEEE conferences occur in a variety of formats. A conference could be very large (with several thousand participants) or rather small (with under a hundred attendees). Most conferences have a single location, however, some are dispersed geographically. Single conference could be even delivered as a series of events separated in time. In this paper, we focus on a conference model, which is arguably the most common one – single location, several concurrent tracks/symposia, several hundred international attendees.

Internet and various web applications are used in virtually all domains of human activities. Preparation and delivery of conferences is one of the examples, where internet is widely used for dissemination of the information about a forthcoming event (through email and websites), collecting paper submissions, registering attendees, and, finally, storing conference papers in the online databases.

Using internet for various types of collaboration has become a mainstream in the industry and academia. Almost any email from a product or service vendor would include an invitation to attend a webinar, and most of us participate in the corporate online meetings on a regular basis.

It is obvious, that similar types of the web conferencing technologies could be used in the delivery of the IEEE conferences. It is just a matter of time when these technologies will be mature enough to meet all the business needs of the conference organizers, and organizing committees will become fully aware of potential benefits and develop skills to incorporate web conferencing technologies in a standard conference procedures.

The purpose of this paper is to draw attention of the conference organizers to the opportunities of the web conferencing, and share the knowledge and experience which have been accumulated during the preparation and delivery of the IEEE Toronto International Conference – Science and Technology for Humanity (TIC-STH) 2009 [1].

## 2.0 Truly Integrated Conference Concept

The use of the web conferencing technologies has been envisioned as a differentiating feature in the delivery of the IEEE TIC-STH 2009 since its inception. The objective of the conference (as it was announced in September 2008) was to deploy the following model: A multi-point worldwide-distributed network of conference authors/participants will enhance the standard (centralized) IEEE conference model, which requires attendance of the participants in person at the main conference location. The participants will be given a choice of delivering conference papers, tutorials, etc. either at the central conference site (hotel) or from their home/office computers wherever they are, eliminating the need of costly and time-consuming travel. This model will require seamless integration of the onsite and online conference systems, including data/presentation, video, audio, feedback, etc. We call this model a “Truly Integrated Conference” (TIC).

Please note that the abbreviation "TIC" also stands for "Toronto International Conference", but in this meaning it is used only as part of a complex abbreviation "IEEE **TIC**-STH 2009".

The main features of the TIC are:

- Use of technology allows onsite and online attendees to have similar opportunities to participate in a conference, have access to information and collaboration.
- Online attendees can deliver their papers live on the internet, as well as watch/listen to the broadcast from the conference site.
- Use of the web conferencing solution which includes web conferencing application integrated with video and audio technology means at the conference site.
- Web conferencing solution provides all of the following channels to facilitate collaboration and make communication “natural”: data/presentation, video, audio, chat.
- The whole conference (not selected parts or tracks) is equally available to onsite and online attendees.

Literature review didn’t reveal any conferences which would utilize a concept similar to the TIC.

Some conferences experimented with web casting of selected sessions (e.g. 2008 IEEE Geoscience and Remote Sensing Symposium [3].

Others, completely moved to the online model (e.g. International Joint Conferences on Computer, Information, and Systems Sciences, and Engineering [4].

Our search has shown, that IEEE TIC-STH 2009 was the first IEEE conference delivered according to the TIC concept.

### 3.0 TIC Concept Benefits And Challenges

***Benefits.*** TIC benefits are multiple and largely quantifiable.

*For Participants.* TIC allows participants to avoid travel to the conference site. An average direct savings per person per conference could be up to one thousand dollars. Also, avoiding travel (especially, international) will “create” a couple of additional productive days for a researcher. An important benefit is increased flexibility. E.g. should the working plans change and keep a person at home, attendee who initially was going to participate onsite, can easily re-register to online participation.

Authors would like to emphasize that they are not promoting online conferences at the “expense” of the traditional onsite events. Nothing could be more stimulating for a researcher than an in-person interaction and exchange of ideas. A significant intangible value of the TIC concept is that it offers participants a freedom of choice – which events to visit in person and in which to participate online.

*For Organizers.* Having certain percentage of participants online, allows organizers to reduce conference food expenses, which could constitute a significant part ( up to 50%) of the registration fee. This statement is based on the assumption that onsite and online attendees pay the same fees. Also, organizers have a good potential to attract many additional attendees without papers who will be only watching the conference online.

*Environmental.* Avoiding travel will have a certain environmental impact. It is easy to estimate that if 30% of the IEEE conference attendees will select the online option, it will result in reduction of millions of pounds of CO2. Some web conferencing providers offer their customers online calculators to assess environmental impact and cost reductions, e.g. http://www.ilinc.com/greenmeter.

***Challenges.*** Using TIC concept to organize a conference requires addressing many challenges.

*Psychological*. As any new endeavour, integrated conferences will have a certain number of enthusiastic supporters. However, there will be even more people either neutral or reluctant to embark on an unknown soil. Some of them just don’t want to take the risks (which certainly exist). Others see TIC as more of an online conference (which is not true), and are concerned that wide spread of this practice would become a roadblock for them to get funding for travel, and limit their opportunities for a face-to-face communication with colleagues.

*Technological.* Selecting an appropriate web conferencing tool which meets the conference requirements is a serious challenge. Incorporating this tool into an onsite audio/video environment could be even more complicated task.

*Organizational.* Documented procedures for web conferencing according to the TIC concept don't exist. It is not only that the technology means must be integrated. Creating a seamlessly integrated conference process requires collaborative efforts from web conferencing and all other conference committees. E.g. such well-established process as paper scheduling needs to be performed taking into account local time of the online attendees, to avoid inconveniences of delivering a paper at 4 AM. Operating web conferencing technology means requires additional human resources, which need to be recruited and trained.

*Financial.* Despite the potential savings for the organizers, web conferencing is not coming for free. TIC implementation requires due diligence in assessing acquisition of services. At a first glance, hourly rates for web conferencing look reasonably low. However, mere multiplication of these numbers by 8 (number of the working hours per conference day) and then by 7 – 10 (number of the conference parallel tracks/symposia), and then yet by 2 – 3 (number of conference days), could shock any treasurer.

*Legal.* IEEE has rigorous and clear policies regarding copyright of conference papers. Paper presentations (PowerPoint slides) are not subject to the same scrutiny as they "exist" for several minutes during the conference and are available to a limited number of attendees in the conference room. Web conferencing actually turns verbal presentations into "documented content" through data, audio, video channels. This content could be delivered to hundreds and thousands people around the world, and potentially could be recorded/stored both at the conference site and by any attendee. Lack of regulation of this issue should be a matter of concern for both organizers and presenters.

All challenges need to be taken into account and addressed by the organizing committee through the best practices of project risk management. Any improper decision may be harmful not only for the web conferencing part, but also for a conference as a whole. Some experiences and recommendations are provided in the IEEE TIC-STH case study below.

## 4.0 TIC Requirements to the Web Conferencing Solution

Selecting a web conferencing application and building a TIC solution for a conference is a complex task. To streamline and facilitate this process we developed a set of requirements presented in Table 1. These requirements address the business needs of both the conference site and equipment of the online attendees.

The following terms are used to describe the requirements.

**Host** - session/meeting chair. Reference to the Host and his/her actions below includes both his/her own actions and actions of the staff assigned to the session.

Other conference support personnel (e.g. Web/AV Coordinator) may be assigned to the Host to operate the session technical means including presentation computer. They will be acting under the control of the Host.

**Presenter** - author of a paper or tutorial presenting his/her work at the conference, who has bee given the rights by the host

Author becomes a presenter when given Presenter's rights by a host.

**Attendee** - person (non-author) who is attending the conference - author who is not presenting his/her paper in the current session.

Some requirements need to be highlighted.

Most important is to identify conference business scenarios which are to be supported by the web conferencing application. Our analysis has shown three types of business scenarios which could be found at most conferences:

- Scenario 1. Keynote/Tutorial/Training Presentation (Figure 1):
  - Broadcast to up to 1,000 simultaneous attendees worldwide
  - Single host at the conference site
  - Presenter may not be located at the conference site
  - Up to Ten (10) tutorial sessions hold simultaneously and independently.

Training scenario refers to the pre-conference educational events aimed at providing organizing committee, symposia TPC and conference technology assisting personnel with knowledge and skills necessary to support web conferencing.

- Scenario 2. Symposium Session (Figure 2):
  - Up to ten (10) presenters per session
  - Presenters are dispersed worldwide
  - Broadcast to up to 1,000 simultaneous attendees
  - Single host at the conference site.

Up to Twenty Five (25) symposia sessions hold simultaneously and independently

- Scenario 3. Conference Discussion Panel (Round Table Session) or Committee Meeting (Figure 3):
  - Up to Fifteen (15) presenters per session
  - Presenters are dispersed worldwide

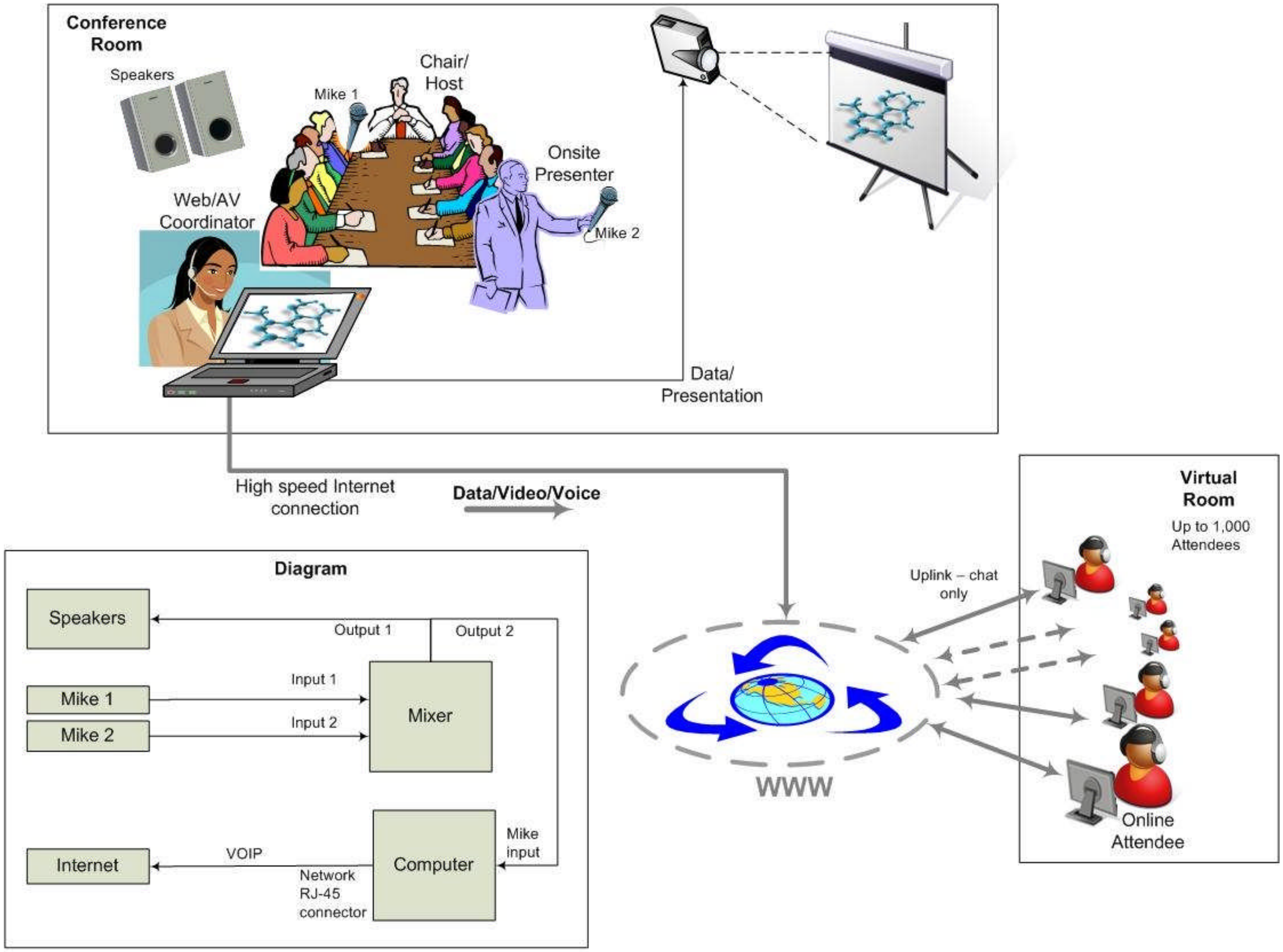

Figure 1: Scenario 1

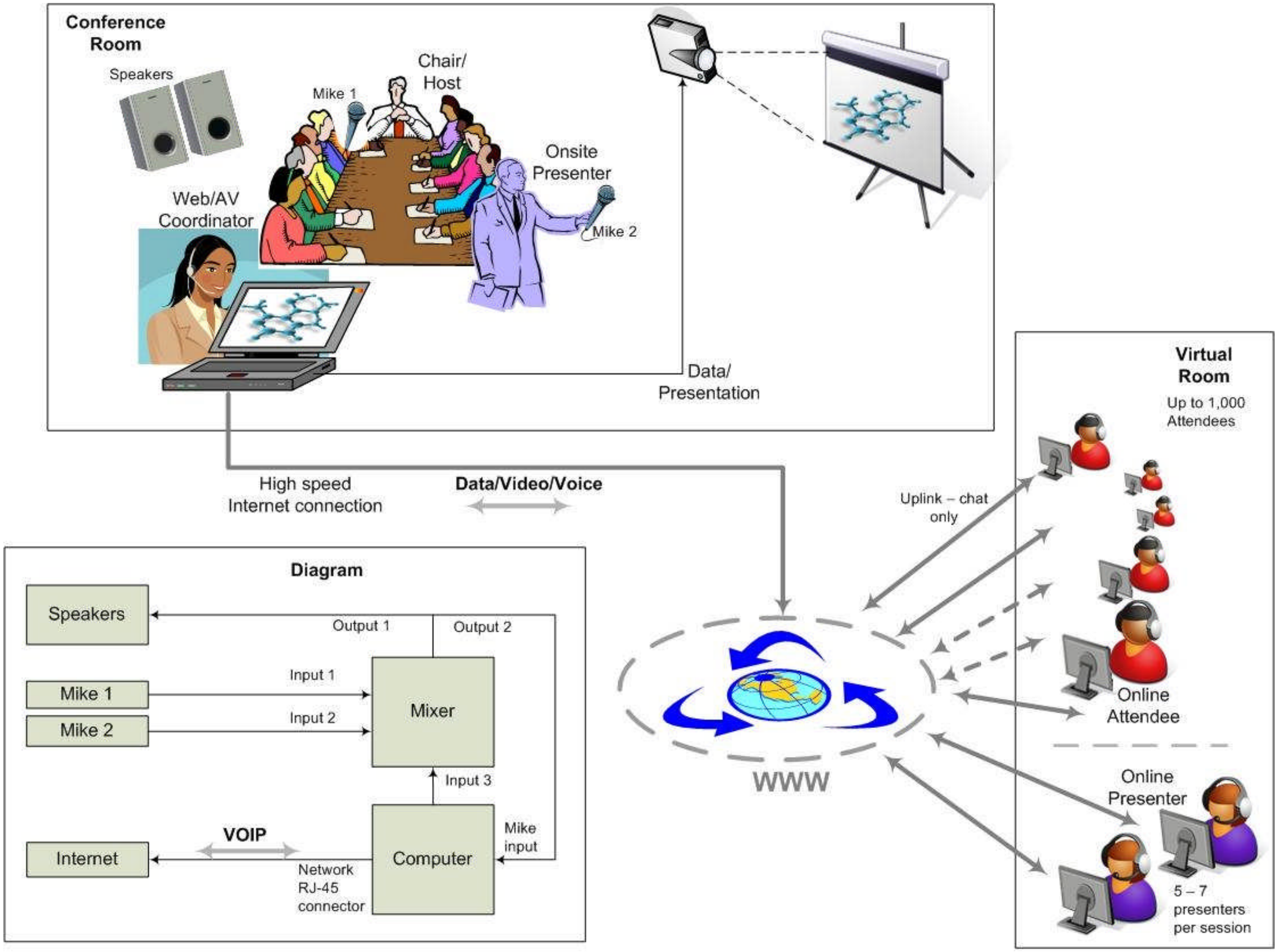

Figure 2: Scenario 2

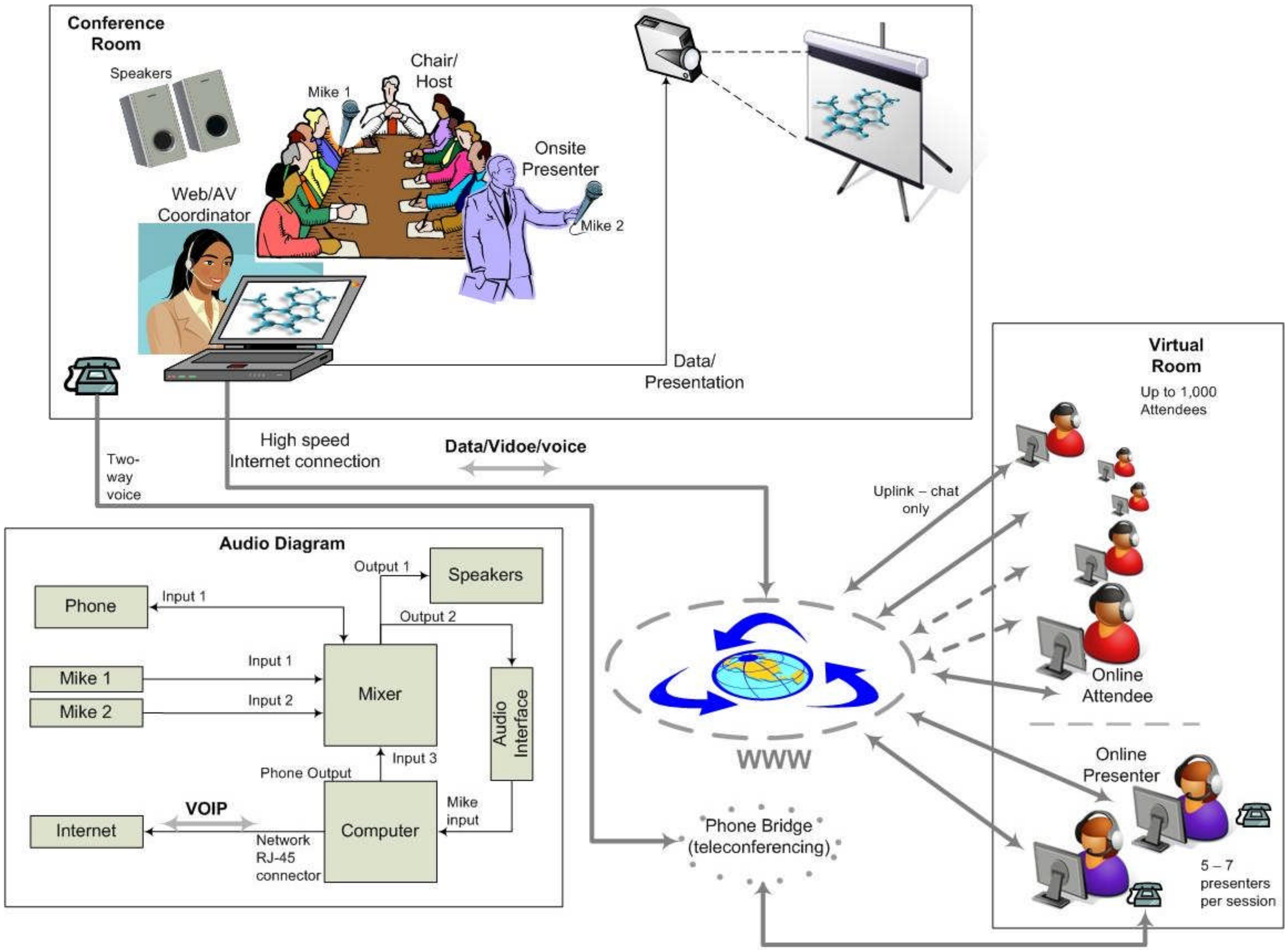

Figure 3: Scenario 3

- Broadcast to up to 200 simultaneous attendees
- Single host at the conference site.

Web conferencing tool shall provide an environment with seamlessly integrated Data/Presentation, Video and Audio channels of communication for the collaboration of the conference participants onsite and online. Although, most of the vendors provide tools which support all three channels, some either don't have a video channel, or don't recommend to activate it in order to save the bandwidth and accommodate higher number of online participants.

Ideally, web conferencing tool should be a truly web based application, which will not require any other software installed on the computers of the online participants (exception could be made for a commonly used software like Flash). This requirement is driven by the intention to make the life of participants easier, avoid possible problems, and also by understanding that some participants who are using corporate computers are often not allowed to install any additional software. The problem with this requirement is that in the "share the screen" mode, which is supported by many tools, installation of additional software/patches is mandatory (although that may not be acknowledged upfront by the vendor). "Share the screen" mode could be very useful, e.g. it could be used for demonstrating any online applications during presentation of papers. Organizers are to make a decision, if they want to limit possible formats of presentations to PowerPoint slides, or to accept risks of allowing screen sharing. That is one of the examples, when decisions made on a web conferencing tool, will affect a conference as a whole (as the onsite participants will also be bound by the decision).

Each conference at the early stages of preparations has to review the suggested requirements based on its own needs, and develop a set of individual requirements to the web conferencing system.

## 5.0 Web Conferencing Applications

According to Wikipedia, web conferencing, also referred to as econference, on-line meeting is used to conduct live meetings, training, or presentations via the Internet. In a web conference, each online participant sits at his or her own computer and is connected to other participants via the internet. This can be either a downloaded application on each of the attendees' computers or a web-based application where the attendees access the meeting by clicking on a link distributed by e-mail (meeting invitation) to enter the conference [11].

A webinar is a term to describe a specific type of web conference. It is typically one-way, from the speaker to the audience with limited audience interaction, such as in a webcast [11].

Considering the target of our web conference, we use the term “web conferencing” as the most appropriate to describe specifically the live or real-time connection of computers to allow a meeting to be held through the exchange of data, video and audio information.

The software for web conferencing has been become increasingly popular in the last few years. In the summer of 1994, there were only two products. Both of them were rather primitive freeware packages. Now, there are well over 60 commercial and freeware products, many of them quite sophisticated, which support conferencing on the web in different kind of forms [14].

A vendor independent testing of the online conferencing software, which is offered as an “application service” over the Internet – without installation of own server, has been conducted and the results are reported in [13]. Table 2 shows comparison of the top 5 web conferencing solutions.

It is concluded in [13] that these top 5 solutions have the following summarized features respectively:

- Citrix GoToMeeting 4.0: user-friendly solution with all the key functions and transparent costs; suitable for Windows and Mac;
- Cisco WebEx: complex solution with a comprehensive set of features, which runs on the most common operating system platforms;
- Netviewer: the solution provides a comprehensive set of functions. In return, subscription fees are high compared with the competition;
- Microsoft Office Live Meeting: the 2007 version stands out as a result of its user-friendly interface and a simplified license model. Can also be used by Macs with restricted functions. It is the only solution among the top 5 that had only integrated VoIP as a teleconferencing service;
- Digital Meeting: user-friendly solution with a comprehensive customer service; recorded meetings can be converted into podcasts.

However, when we looked for an appropriate web conferencing solution for IEEE TIC-STH 2009, we could not simply select one of the solutions from the above top ranking list. To make a choice, we had to consider operating system support, compatibility with other environments, administrative capabilities, browser support, customizability, and, certainly, price.

During preparations of the IEEE TIC-STH 2009, web conferencing committee reviewed many commercial tools against the conference requirements and conducted trials of several applications including WebEx, GoToWebinar, Adobe Connect, iVocalize, Instant Presenter.

One of the observations is that there are two types of solutions supporting two models:
- Webinar model - central broadcasting to hundreds and thousand users e.g. GoToWebinar;
- Corporate meeting model (multi-point participants), but the number of participants is limited to 15 - 20, e.g. WebEx.

Both GoToWebinar and WebEx tools were very easy to set up, and no problems occurred during the trials. However, GoToWebinar was not selected for two reasons. First, it doesn't have video. Second, it requires installation of a software patch, which is not allowed in some corporate environments. WebEx was not selected because the cost projections (for the scenarios with several hundred participants) turned out to be prohibitive for the conference budget.

An interesting solution of combined use of the GoToWebinar and WebEx has been proposed and tested that overcomes main problems of the individual tools mentioned above. To realize the solution, several (up to 15 - 20) participants make connection using WebEx, and then one of these participants turns on GoToWebinar in a screen sharing mode and starts broadcast of his screen (with WebEx application running with slides and video) to hundreds of online attendees. Trials confirmed that WebEx video channel was broadcast by GoToWebinar without a problem. So, this solution enriches GoToWebinar with a video channel and at the same time allows WebEx to be received by hundreds of participants at a low cost. This solution could be helpful to deliver web conferencing according to the Scenario 3.

Audio channel. Most vendors provide a choice of voice over internet protocol (VOIP) or teleconferencing. It's revealing that usually the use of a phone line and a conference bridge is the recommended option, primarily for the sound quality reasons. The use of the phone to support a multi-hour conference with participants around the globe, will most likely become unfeasible for budgetary reasons as phone channel is not included in the price of the web conferencing services. Another solution - streaming video and audio - provides an excellent quality of the audio channel. ON24 [8] is an example. However, prices for the streaming video and audio are usually very high.

Video channel. Usually worked well for all products. It was noted that lighting of the presenter sitting in front of the camera is as important as the quality of the camera.

The conclusion was made that none of the commercially available web conferencing solutions fully satisfies the TIC concept requirements. Organizers have to make trade-offs between functionality, quality and costs based on the individual conference requirements.

Based on the above considerations and IEEE TIC-STH 2009 conference requirements, ePresence [9] was selected as a web conferencing system.

ePresence is an Open Source web conferencing and webcasting system that delivers live or on-demand rich media over the internet. Developed by University of Toronto, ePresence is a complete solution for streaming, capturing, and publishing rich media presentations. Considering ease of technical support, price, availability of ePresence Capture Stations offered by our conference host, Ryerson University, ePresence was eventually selected as a foundation of the TIC solution. In next section, more details of the TIC implementation are presented as a case study analysis.

**6.0 Case Study: IEEE TIC-STH 2009**

IEEE TIC-STH 2009 was organized and hosted by the IEEE Toronto Section [2]. Ryerson University was used as a venue. The conference was focused on advanced interdisciplinary problems across a broad spectrum of the IEEE fields of interest. The scope was not limited to the traditional IEEE areas – electrical, computing, and engineering. There were very strong papers in education, social implications of technology and sustainable development of the society.

The conference attracted 360 papers out of which 186 papers were accepted after rigorous peer-review process (~50% acceptance rate). Authors represented 29 countries – see Figure 4: Conference Authors by Region.

IEEE TIC-STH 2009 turned out to be a huge success story from all perspectives: technical, organizational, financial and customer satisfaction. Realization of the TIC concept was one of the drivers for this success.

IEEE TIC-STH 2009 was the first truly integrated conference that realized the scenarios 1 and 2 (described in the Requirements Section) for all the scheduled presentations. The facilitation of the online platform was done by ePresence conference services. There were at maximum 8 concurrent (onsite/online) presentations to be handled at the same time. For this purpose 16 volunteers were trained to work with the conference web conferencing system. Each web conferencing volunteer received 10 hours of training and had to pass the exam to be able to contribute effectively to IEEE TIC-STH in web conferencing. This greatly helped to run the online sessions more smoothly during the conference. Our web conferencing platform was run on the ePresence hosting service [10]. Online users (attendees/presenters) needed to register beforehand to be able to attend/present. The accessibility level of all the online sessions was restricted to our registered users who had paid fully for the registration fee of IEEE TIC-STH 2009.

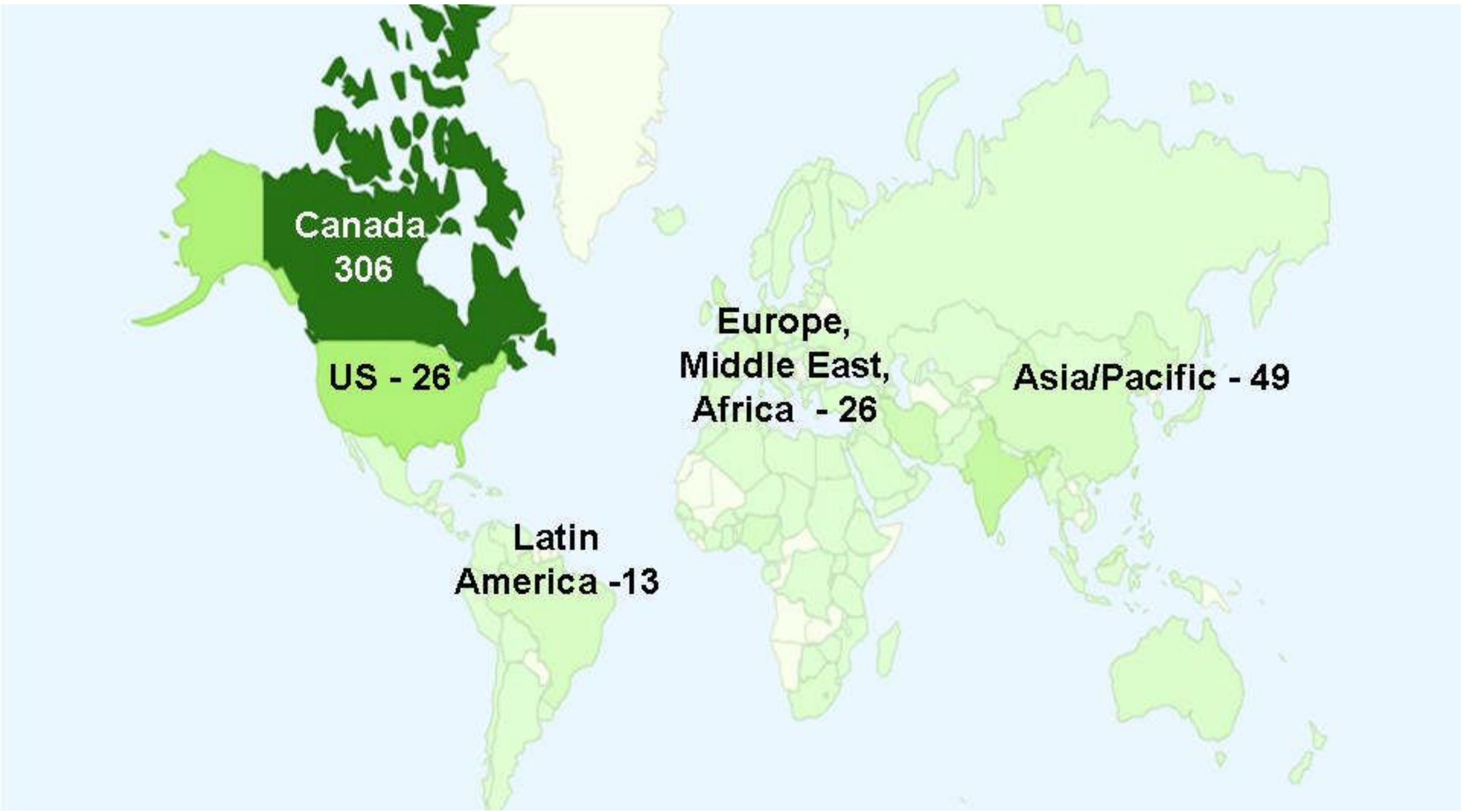

Figure 4: Conference Authors by Region.

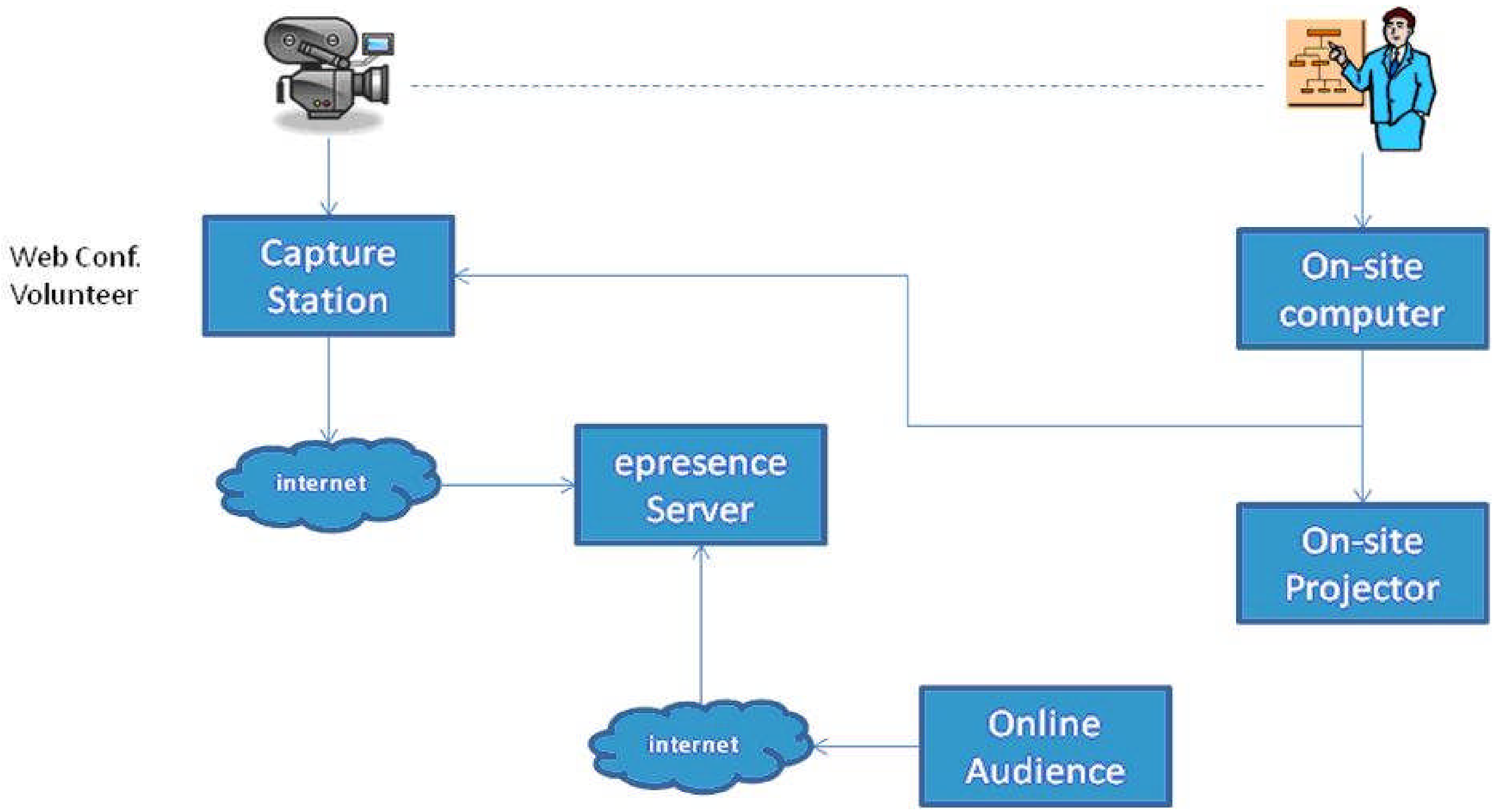

Figure 5: Realization of scenario 1, onsite presentation and online attendee

Scenario 1 was utilized by ePresence capture stations integrated with the presentation system in each presentation room as is shown in the Figure 5.
To assist the online attendees, an instruction manual was produced and uploaded on YouTube [5]. This video manual describes all the required steps for the online attendee to be able to follow a presentation online.

All the concurrent presentations were broadcasted online for our online attendees and also for our onsite attendees who were interested to follow other concurrent presentations online while attending an onsite presentation. This provides the possibility of switching between presentations by a click and attending the presentation that fits best within the interest of the attendee. Online attendees, similar to the onsite attendees could ask questions using chat functionality.
Scenario 2 was utilized by remote presenter connecting to the ePresence server as shown in Figure 6.
The volunteers in the presentation rooms were advised to screen the online presentation. The coordination of the presentations with online presenter was done by means of chatting in the ePresence platform and backed up by a phone connection.

Keeping the conference completely broadcasted online takes a lot of effort in troubleshooting and management. To make sure that the troubleshooting process would be handled as quickly as possible a separate group was hired. The common observed problems before/during the conference were as follows:

1. There is a lot of hardware involved to capture/broadcast all the presentations and there is a high chance that camera, microphone, computer, projector and capture station might not work properly.
2. The high speed connection with the web conferencing server needs to be monitored and maintained continuously.
3. The online presenters need to be connected to a reliable high-speed connection.
4. The presentations should be scheduled considering the fact that the online presenters have to present their work according to their local time.

In IEEE TIC-STH 2009, there were altogether 28 papers that were presented remotely out of almost 200 conference papers. The rest of the papers were presented onsite. All the papers were presented onsite or online and there was no paper which was not presented due to the fact that the authors could not be present at the location of the conference. The web conferencing website was visited 1300 times for following the presentations. Another possible benefit of web conferencing-enabled conference is that not only all the presentations would be broadcasted online but also all the presentations can be recorded for later publication. The recorded presentations can be even bundled with the paper text and be made accessible on IEEE Xplore website. The possibility of watching a presentation of the paper makes it easier for the reader to understand the paper when accessed later on the database. These benefits are supported by the existing technology, however certain legal and organizational issues need to be resolved first.
Although the web conferencing tool is the backbone of the TIC concept, there are other applications that will help organizers.

- Google Analytics [6] was used by IEEE TIC-STH 2009 to collect and analyze conference website traffic, and understand what types of operating systems and web browsers are used by prospective participants. The results of this analysis were included in the requirements to the web conferencing tool.

- Large files transfer system FilesDirect [7] was used to send conference materials to the online attendees.

Post-conference attendee survey shows good customer experience of the online attendees. Although the conference was not immune to some roughnesses, the overall level of satisfaction of the online audience is even higher than that of the onsite participants - see Figure 7 (Participants were responding to the statement/question “Overall – the conference was well organized”).

**7.0 Conclusions and Recommendations**

1. The concept of the Truly Integrated (onsite and online) Conference (TIC) has been suggested and discussed. A conference delivered according to the TIC concept presents significant benefits to both conference organizers and participants. Certain environmental impact also should be assessed and considered.

2. Authors are confident that conferences of this type will soon become common for the IEEE and other organizations.

3. Business and technical requirements to the web conferencing system, satisfying the TIC concept, and its

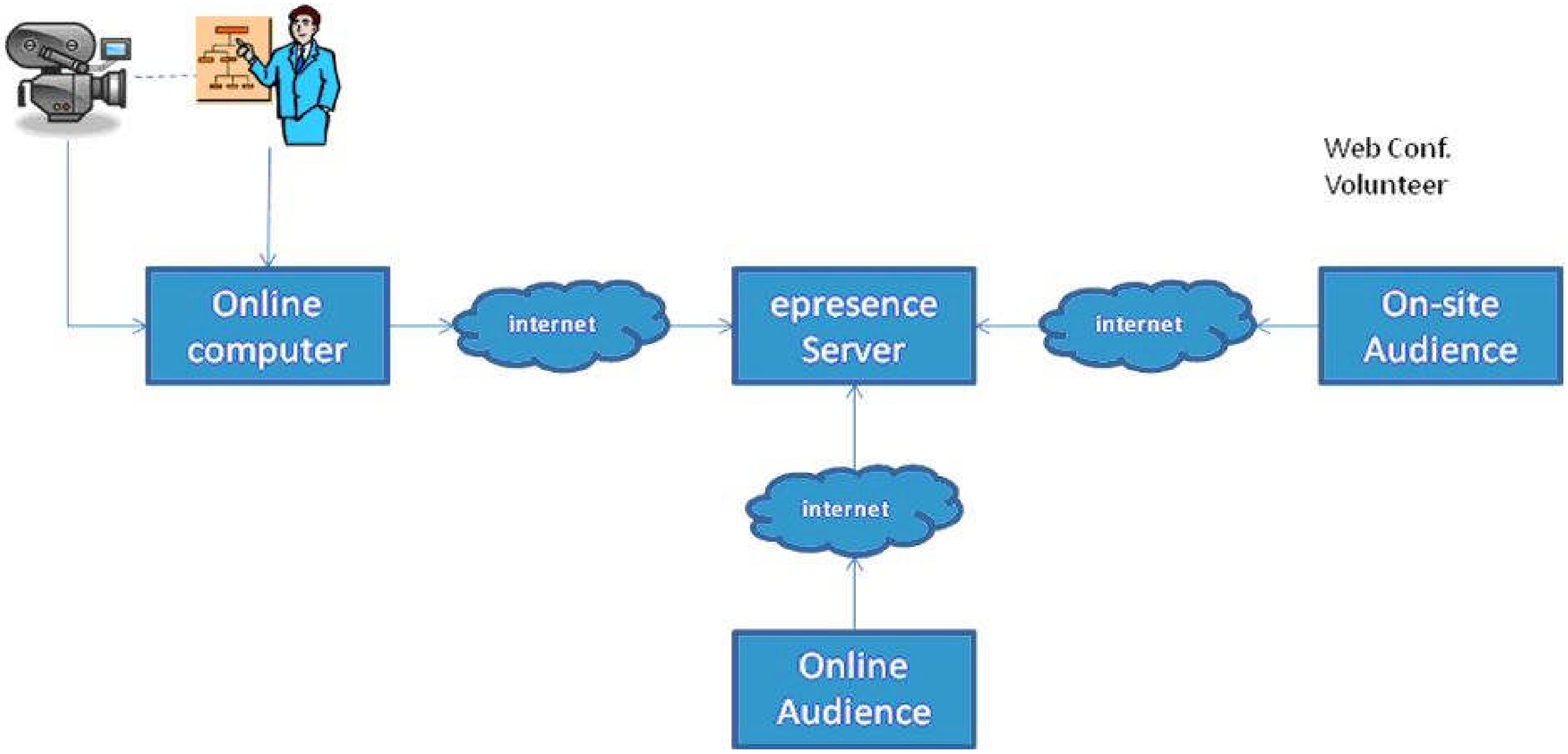

Figure 6: Realization of Scenario 2, online presenter, online/onsite attendee

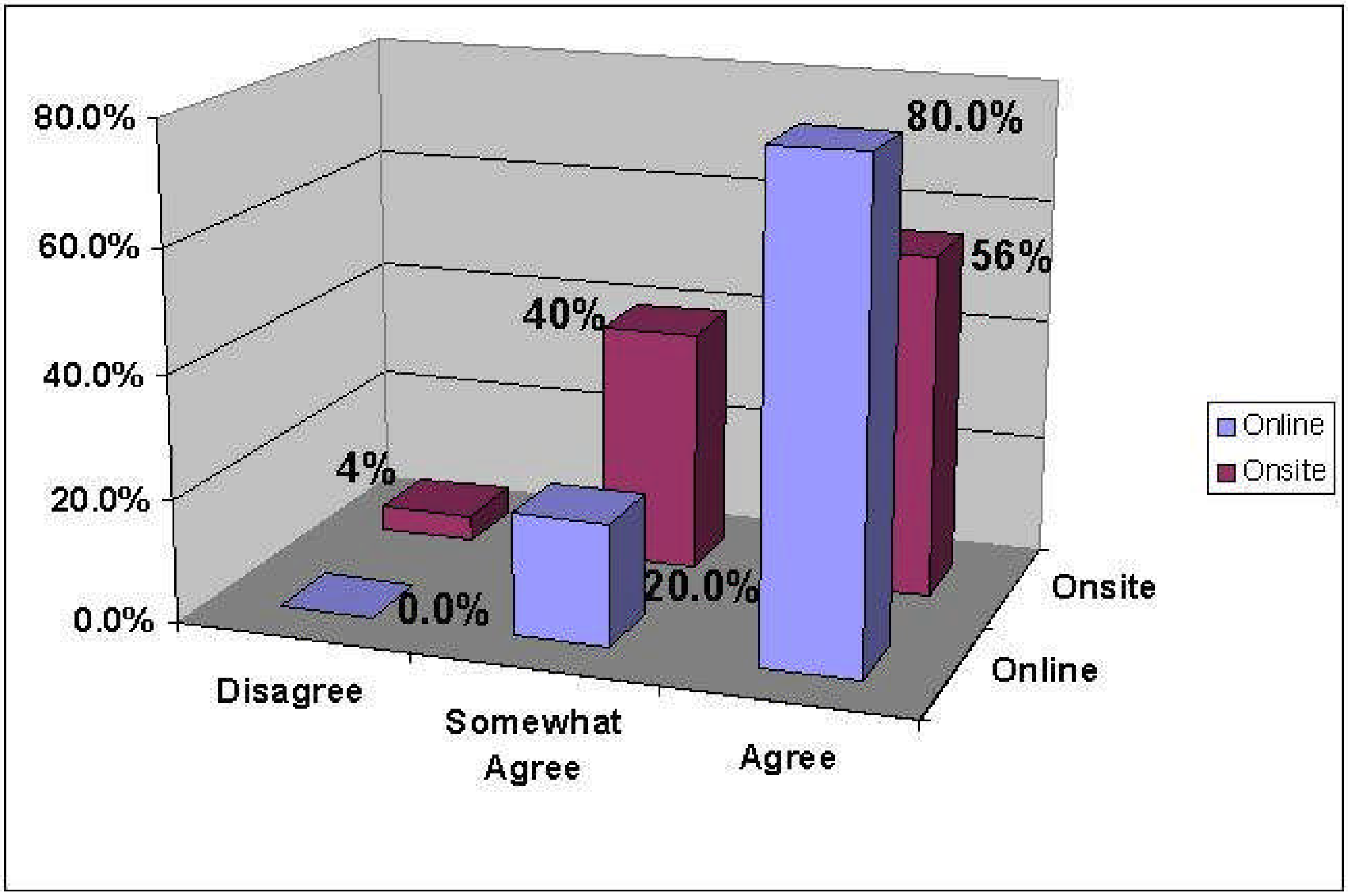

Figure 7: Participants' feedback (Overall – the conference well organized)

incorporation into an integrated solution have been developed, and tested.

4. Analysis of the commercially available web conferencing solutions has shown that none of them fully satisfies the TIC concept requirements. Conference organizers have to prioritize requirements based on the specific needs of the conference and, most likely, give up on less important ones.

5. Our search shows, that IEEE Toronto International Conference - Science and Technology for Humanity (TIC-STH) 2009, organized by the IEEE Toronto Section, was the first IEEE (and maybe just the first) conference delivered according to the TIC concept.

6. IEEE TIC-STH 2009 successfully used ePresence web conferencing application.

## About the Authors

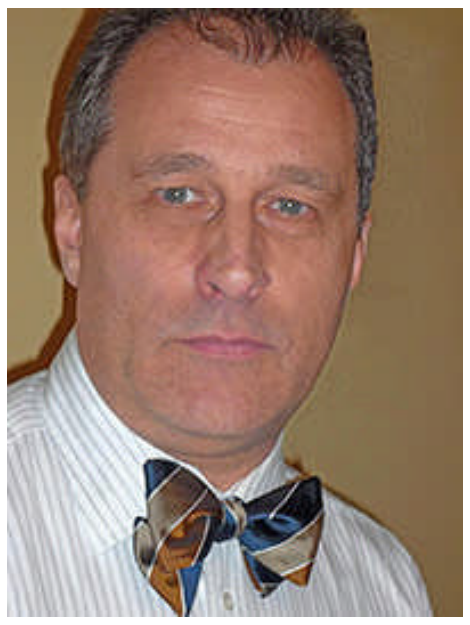

**Alexei Botchkarev** is an information and knowledge management professional, consultant and researcher (www.gsrc.ca). He has over 30 years of experience in project management, systems analysis, modelling and simulation, business processes analysis, information systems solutions, requirements analysis and scientific research. He holds B.Eng. 5-year degree from the Kiev Aviation Engineering Academy, Ukraine (1975) and Ph.D. from the aerospace R&D Institute (1985). He is currently is a Senior Information Management Advisor at the Knowledge Management Branch, Ministry of Health and Long-Term Care, Ontario, Canada. In the course of his career, Dr. Botchkarev worked as an employee or consultant in multiple industries (aerospace, information technologies, advanced materials, healthcare, education and training) and in various capacities (program/project manager, product manager, marketing manager, research analyst) in Canada, US, Europe, and Asia. He has more than 15 years of successful career in the aerospace R&D and manufacturing environment with expertise in managing/coordinating R&D projects. He has solid background in the quantitative and qualitative research methodologies, and simulation modelling of complex processes and systems. 7 years of patent research and coordinating intellectual property issues. 5 years in the education industry in Toronto. His current research interests are in information and knowledge management methodology, enterprise and business architecture, modelling and simulation, quantitative methods in project management and performance evaluation. He has been Chair of the IEEE Toronto Section and a member of the IEEE Canada Board of Directors since 2008. He was founding member of the IEEE TIC-STH 2009 Organizing Committee.

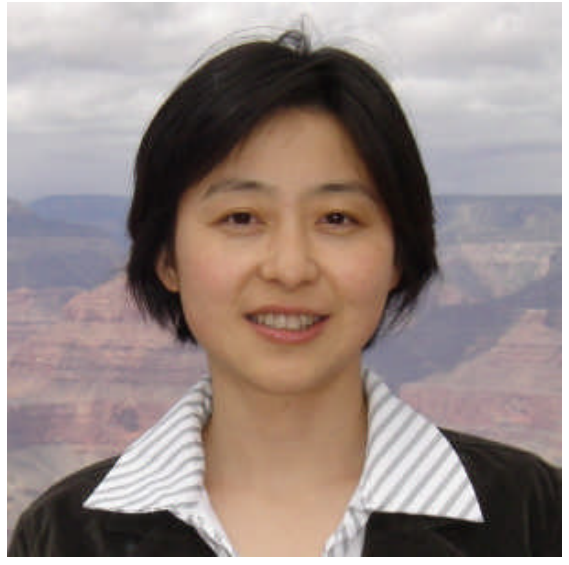

**Lian Zhao** received the Ph.D. degree in electrical and computer engineering from the University of Waterloo, Canada, in 2002. Before she joined Ryerson University in 2003, she worked as a postdoctoral fellow with the Center for Wireless Communications, University of Waterloo. She is an Associate Professor at the Department of Electrical and Computer Engineering, Ryerson University, Toronto, Canada. She is a cofounder of the Optic Fiber Sensing Wireless Network Laboratory in 2004. Her research interests are in the areas of wireless communications, radio resource management, power control, as well as design and applications of the energy efficient wireless sensor networks. She is an IEEE Senior Member and a Registered Professional Engineer in the province of Ontario, Canada. Lian Zhao was Chair, Web Conferencing Committee of the IEEE TIC-STH 2009.

**Hamed Rasouli** is a Ph.D. candidate in Radio Resource Management and Radio Access and Networking (RRM+RAN) research group at the Department of Electrical and Computer Engineering, Ryerson University, Toronto.

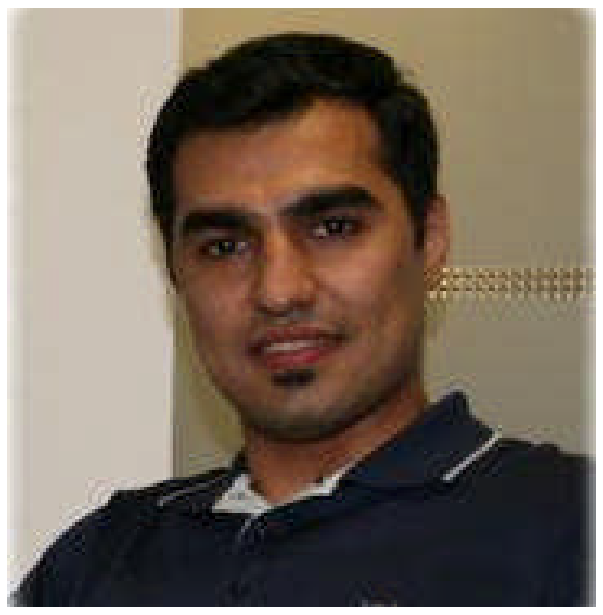

He received his B.Sc. with Distinction in Electrical Engineering – Telecommunications from University of Tehran, Iran in 2003 and completed his M.Sc. at Iran University of Science and Technology afterwards. His research interests include radio resource allocation and management in cooperative and cognitive radio wireless communication systems. Hamed Rasouli was Vice Chair, Web Conferencing Committee of the IEEE TIC-STH 2009.

**Table 1.** Web Conferencing Solution Requirements

| **No** | **Requirement** | **Comment/ Description** | **Importance (Must Have or Nice to Have)** |
|---|---|---|---|
| | | | |
| | **Web Conferencing Software/Service** | | |
| 1 | Shall provide an environment with seamlessly integrated Data, Video and Audio channels of communication for the collaboration of the conference participants onsite and online according to the **"Truly Integrated Conference" (TIC)** concept. | | Must Have |
| 2 | Shall be a single solution capable of supporting all types of the conference business scenarios: | | Must Have |
| 2.1 | Scenario 1. Keynote/Tutorial/Training Presentation (Figure 1) | | Must Have |
| 2.2 | Scenario 2. Symposium Session (Figure 2) | | Must Have |
| 2.3 | Scenario 3. Conference Discussion Panel (Round Table Session) or Committee Meeting (Figure 3) | | Nice to Have |
| 3 | Shall allow a host to select and grant /withdraw Presenter's rights | | Must Have |
| 4 | Shall allow Presenters to: | | |
| 4.1 | • Conduct slide presentations using: | | |
| 4.1.1 | - PowerPoint slides saved in PDF format | Some features of the PowerPoint slides are distorted by web conferencing (colors, shades, etc.) | Must Have |
| 4.1.2 | - PowerPoint slides | | Must Have |
| 4.1.3 | - Mac Keynote or other presentation software | | Nice to Have |
| 4.2 | • Conduct presentations using desktop sharing in real time | | Nice to Have |
| 4.3 | • While presenting, highlight information on the slides to emphasize/clarify presentation | | Nice to Have |
| 4.4 | • Display live video using a Webcam or digital camera | | Must Have |
| 4.5 | • Have two-way interactive audio using microphone and headset | | Must Have |
| 5 | Shall allow recording (video and audio) of the sessions by the Host | | Must Have |
| 6 | Shall deny opportunity of recording (video and audio) by the attendees/presenters | | Nice to Have |
| 7 | Shall allow means for conducting surveys of the conference participants in real time | | Must Have |
| 8 | Record feedback and individual responses | | Nice to Have |
| 9 | Shall allow customization of the conference webpage viewed by attendees (conference logo) | | Must Have |
| 10 | Shall allow all attendees to submit questions to Presenters through the Host using Chat feature | | Must Have |
| 11 | Shall support browsers: | | |
| 11.1 | Internet Explorer | 50% of potential attendees | Must Have |
| 11.2 | Firefox | 40% of potential attendees | Must Have |
| 11.3 | Safari, Chrome | 7% of potential attendees | Nice to Have |
| 12 | Shall support operating systems: | | |
| 12.3 | Windows | 88% of potential attendees | Must Have |
| 12.4 | Macintosh | 7% of potential attendees | Must Have |
| 12.5 | Linux | 4% of potential attendees | Nice to Have |
| 13 | Shall provide opportunity for pre- registration of attendees | | Must Have |
| 14 | Shall have advanced security mechanisms for authorized password protected access to the sessions | | Must Have |
| 15 | Shall provide opportunity for controlling and recording individual attendance, and reporting features (session, attendee name, time in and out) | E.G. must have this info for sessions where attendees will be granted PMI's PDUs<br>Also, need this for statistics. | Must Have |
| 16 | Shall have advanced security mechanisms to keep all information during and after sessions protected from unauthorized access | | Must Have |
| 17 | Web conferencing software shall be available as a service (SaaS) | No on-premise software | Must Have |
| 18 | Conferencing solution shall be web based | | Must Have |
| 19 | Using conferencing solution shall not require installation of any additional software (add-ins) on the attendees/presenters' computers | Except widely used generic purpose software, e.g. Flash | Must Have |
| 20 | Using conferencing solution shall not require firewall reconfiguration, even for remote desktop sharing | Advanced firewall compatibility | Must Have |
| 21 | Connecting to the web conferencing solution for a scheduled session shall require only:<br>- URL of the session<br>- Attendee's name<br>- Attendee's password | | Must Have |
| 22 | Shall provide features for testing compatibility of the attendees' and presenters' computers with web conferencing software/service | | Must Have |
| 23 | Web conferencing Vendor shall provide technical support to attendees | | Nice to Have |
| 24 | Vendor shall provide training to attendees | | Nice to Have |
| 25 | Vendor shall guarantee System Reliability and Redundancy | | Nice to Have |
| 26 | Shall allow multiple concurrent sessions | | Must Have |
| | **Conference Site** | | |
| 27 | High speed Internet connection | | |
| 28 | Computer: | | |
| 28.1 | • Intel or AMD processor (1GHz or faster) | | |
| 28.2 | - At least 512 MB RAM (at least 2 GB RAM for Vista) | | |
| 28.3 | • Windows 2000, XP, 2003 and Vista<br>• Or MacOSX 10.4,10.5 | | |
| 28.4 | • Internet Explorer 6/7/8<br>• or Firefox 2/3 | | |
| 29 | Projector | | |
| 30 | Telephone/Phone line | | |
| 31 | Microphones Wireless (2) | e.g. Shure SM58-LC | |
| 32 | Speakers (2) with stands | e.g. Yorkville E10P | |
| 33 | Mixer | e.g. MACKIE 12CH | |
| 34 | Audio Interface | e.g. FastTrack Pro | |
| 35 | Lighting | Lighting is important for video quality | |
| | **Participants (Presenters/Attendees)** | | |
| 36 | Internet connection | High speed connection recommended | |
| 37 | Computer: | | |
| 37.1 | • Intel or AMD processor (1GHz or faster) | | |
| 37.1.1 | o At least 512 MB RAM (at least 2 GB RAM for Vista) | | |
| 37.1.2 | o Windows 2000, XP, 2003 and Vista<br>o Or MacOSX 10.4,10.5 | | |
| 37.2 | • Internet Explorer 6/7/8<br>• or Firefox 2/3 |  |  |
| 38 | Telephone/Phone line |  |  |
| 39 | Microphone/speakers | Headset recommended |  |
| 40 | Web camera | For presenters only |  |
| 41 | Lighting device | For presenters only |  |

**Table 2: Web Conferencing Applications Ratings**

| No | Solution | Number and Scope of Features | User-friendliness | Meeting setup effort | Software installation effort | Pricing / cost transparency | Security | System requirement | Overall score |
|---|---|---|---|---|---|---|---|---|---|
| 1 | Citrix GoToMeeting 4.0 | **4** | **5** | **5** | **5** | **5** | **4** | **4** | **9.49** |
| 2 | Cisco WebEx | **4** | **3** | **5** | **5** | **5** | **4** | **5** | **9.26** |
| 3 | Netviewer | **5** | **4** | **5** | **5** | **3** | **4** | **3** | **9.15** |
| 4 | Microsoft Office Live Meeting 2007 | **4** | **2** | **5** | **2** | **5** | **4** | **3** | **8.76** |
| 5 | Digital Meeting | **3** | **5** | **4** | **5** | **5** | **2** | **4** | **8.74** |